\documentclass[12pt,psfig]{article}
\usepackage{amssymb}
\usepackage{amsmath,amsthm}
\usepackage{amsfonts}
\usepackage{graphicx}
\usepackage{graphics}
\usepackage{indentfirst}
\usepackage{color,graphicx}
\usepackage{caption}
\usepackage[utf8]{inputenc} 
\usepackage[english]{babel}
\usepackage[nottoc]{tocbibind}
\usepackage[usenames,dvipsnames]{xcolor}
\usepackage[symbol]{footmisc}
\numberwithin{equation}{section}

\usepackage{hyperref}

\begin{document}
\begin{center}\Large\textbf{The Inflationary 
Dynamics with the Scalar-Tensor Model}
\end{center}
\vspace{0.75cm}
\begin{center}{\large Feyzollah Younesizadeh 
and \large Davoud
Kamani} {\footnote{\textcolor{Magenta}
{Corresponding author}}}
\end{center}
\begin{center}
\textsl{\small{Department of Physics, 
Amirkabir University of
Technology (Tehran Polytechnic) \\
P.O.Box: 15875-4413, Tehran, Iran \\
e-mails: fyounesizadeh@aut.ac.ir , 
kamani@aut.ac.ir \\}}
\end{center}
\vspace{0.5cm}

\begin{abstract}

We investigate the cosmic inflation within a class of the 
scalar-tensor model with the scalar-dependent 
non-minimal kinetic couplings. The inflationary dynamical 
potential will be applied. 
Using the slow-roll approximation, 
we compute theoretical predictions for the key observables, 
like the spectral indexes $n_s$, scalar-to-tensor ratio $r$ 
and the running of the scalar spectral index $\alpha_s$ in 
terms of the free parameters of the model. Besides, we 
find the limitations of these parameters. In addition, these 
quantities will be compared with the latest observational 
data from the Planck data. Furthermore, we analyze the 
sensitivity of $r$, $n_s$ and $\alpha_s$ in terms of 
the model's free parameters.

\end{abstract}


\newpage
\section{Introduction}

The cosmological inflation explains the
present large-scale structure. Besides,
it addresses the important problems in the standard
cosmology \cite{1}-\cite{8}. 
In extensive studies of the inflationary models,
many theories of the gravity have been recently taken 
into consideration \cite{9}-\cite{36}.
A major challenge in constructing the inflationary models 
is the flatness and stability of the inflaton potential
against the quantum corrections 
to sustain the slow-roll inflation. 
The quantum fluctuations during the inflation act 
as the initial conditions for the large-scale 
structures and anisotropies \cite{23}-\cite{25}. 

We consider a special theoretical 
framework for investigating the inflation.
It is called dynamical inflation (DI), that solves 
certain limitations inherent in the conventional 
inflationary models, including the adjusting 
of the inflation potential and the initial 
conditions for it \cite{5}-\cite{8}, \cite{37}, \cite{38}. 

In fact, the DI potential fails to achieve a successful 
transition out of the inflation and results in the 
spectral index $n_s$ that exceeds the current 
observational thresholds. By incorporating further 
interactions, such as the non-minimal kinetic 
coupling, we can improve the limitations of 
this potential. Hence, the spectral index $n_s$ 
can be adjusted to align with the most recent 
observational data \cite{27}-\cite{29}, which 
also facilitates a graceful exit from 
the inflationary phase.

The primary motivation for this work is to construct a 
robust and well-motivated extension of the DI scenario 
that can successfully align with the modern cosmological 
data. While the standard DI framework is theoretically 
appealing, its failure to produce a scalar spectral index 
$n_s$ within the Planck bounds and its challenges 
with a graceful exit from inflation 
\cite{37}, \cite{38}, necessitate a physical 
modification rather than an ad hoc adjustment of the 
potential. Therefore, we introduce a non-minimal kinetic 
coupling of the form $J(\phi) G_{\mu\nu} \partial^
\mu \phi \partial^\nu \phi$. This choice is strongly 
motivated by a high-energy physics perspective.
In fact, such couplings naturally arise 
in the low-energy 
effective action of the string theory and 
they are part of the 
healthy, the ghost-free Horndeski class of scalar-tensor 
theories \cite{39}, \cite{40}. This provides a compelling 
and theoretically 
consistent avenue for modifying the inflationary dynamics.

This model offers some specific advantages 
relative to the standard 
slow-roll inflation in the general relativity(GR). 
Firstly, the coupling $J(\phi)$ effectively flattens 
the scalar field potential during the inflation, not by 
fine-tuning $V(\phi)$ itself but by modifying 
the kinetic energy of the inflaton. This naturally 
suppresses the tensor-to-scalar ratio $r$ and it can 
significantly adjust the scalar spectral index $n_s$, 
pulling it into the observationally favored range. 
Secondly, this interaction 
can facilitate a graceful exit from the inflationary phase. 
The key limitation, common to many beyond-GR models, 
is the introduction of new functional degrees of freedom.
Here, it is the functional $J(\phi)$ with its parameters 
$\alpha, \beta$ and $\mu$. Its possible naturalness 
issue at high energies is the largeness of $J(\phi)$. 
However, as we shall demonstrate, 
these parameters are not arbitrarily fine-tuned.
In fact, they are constrained to specific natural ranges by 
the observational data, 
and their impact is physically interpretable through the 
model's sensitivity. 

In this paper, we investigate the inflation
via a modified gravity, which is 
characterized by the kinematically coupling term  
$J(\phi)G_{\mu\nu}\partial^{\mu}\phi\partial^{\nu}\phi$. 
We shall apply a special DI potential and an 
appropriate form for the functional $J(\phi)$,
which includes some free parameters. 
Within the slow-roll approximation, we derive 
precise analytical expressions for the key physical 
quantities such as 
the scalar spectral index $n_s$, the 
tensor-to-scalar ratio $r$ and the running of 
the scalar spectral index $\alpha_s$. 
According to the Planck data, we observe that
this model leads to a more 
successful inflation than the standard DI model. 
Besides, we present sensitivity plots of the 
$n_s$, $r$ and $\alpha_s$ in terms of the foregoing free 
parameters of the model.

This paper is organized as follows. 
In Sec. \ref{200}, we derive the equations 
of motion and the slow-roll parameters 
``$n_s$'', ``$r$'' and ``$\alpha_s$''. In Sec. \ref{300}, 
the DI via our model will be investigated.
This provides a comparative 
analysis of our findings against the data from 
the Planck 2018. Finally, the section 
\ref{400} is devoted to the results and conclusions.

\section{The setup and field equations}
\label{200}

We consider a generalized scalar-tensor 
model with a non-minimal 
kinetic coupling of the real scalar field 
$\phi$ to the Einstein's 
tensor. In the Einstein frame, the corresponding action 
can be written as, \footnote{We use the natural units
$c =\hbar = 1$, $\kappa^2=M^{-2}_p=8\pi G$.}
\begin{equation}
\label{2.1}
S=\int{{\rm d}^4x\sqrt{-g}}\left(\frac
{1}{2\kappa^2}R-
\frac{1}{2}\partial_{\mu}
\phi \partial^{\mu}\phi-V(\phi)+
J(\phi)G_{\mu\nu}\partial^
{\mu}\phi\partial^{\nu}\phi\right),
\end{equation}
where $G_{\mu\nu}=R_{\mu\nu}-\frac{1}{2}g_{\mu\nu}R$ 
is the Einstein's tensor. The functional $J(\phi)$ is  
differentiable, and $V(\phi)$ is the potential 
of the scalar field. For the derivations in this 
section, we work in units where $M_P = 1$. 
However, we shall restore it 
in the section 3 for consistency with observational data.

We assume a homogeneous scalar field $\phi=\phi(t)$, 
in a homogeneous and isotropic FRW background metric
\begin{equation}
\label{2.2}
ds^2=-dt^2+a(t)^2(dx^2+dy^2+dz^2).
\end{equation}
Thus, the corresponding equations 
of motion find the following features 
\begin{equation}
\label{2.3}
3H^2=\frac{1}{2}\dot\phi^2+V(\phi)+
9H^2J(\phi)\dot\phi^2,
\end{equation}
\begin{equation}
\label{2.4}
-2\dot H=\dot\phi^2\bigg(1+
6H^2J(\phi)\bigg)-
2\frac{d}{dt}\bigg(HJ(\phi)\dot\phi^2\bigg),
\end{equation}
\begin{equation}
\label{2.5}
\ddot\phi+3H\dot\phi+V^{\prime}
(\phi)=-6HJ(\phi)
\dot\phi\left(3H^2+2\dot H\right)
-3H^2\left(2J(\phi)\ddot\phi+J^{\prime}(\phi)
\dot\phi^2\right).
\end{equation}
The Hubble parameter (function) is defined by  
$H\equiv \dot a(t)/a(t)$. For the special case 
$J = 0$, our model reduces to the 
standard Friedman equations with the single 
scalar field. 

Through a straightforward calculation, we derive 
the following expressions for $\dot \phi^2$ and $V(\phi)$ 
from the foregoing field equations
\begin{equation}
\label{2.6}
V(\phi)=H^2\left(3-\epsilon_1-2
\ell_0-\frac{1}{3}\ell_0 (\ell_1-\epsilon_1)\right),
\end{equation}
\begin{equation}
\label{2.7}
\dot\phi^2=H^2\left(2\epsilon_1 -2\ell_0+\frac{2}{3}\ell_0
(\ell_1-\epsilon_1)\right),
\end{equation}
where the parameters, which are specifically characterized 
during the inflationary phase, are given by
\begin{equation}
\label{2.8}
\epsilon_1= -\frac{\dot{H}}{H^2}, \quad \epsilon_2
=\eta= \frac{\dot{\epsilon_1}}{H\epsilon_1}, \quad 
\epsilon_3= \frac{\dot{\epsilon_2}}{H\epsilon_2}, 
\quad \ell_0=
3J(\phi)\dot \phi^2, \quad \ell_1=
\frac{\dot \ell_0 }{H\ell_0}.
\end{equation}
where $\epsilon_1$ and $\epsilon_2$ 
represent the Hubble flow 
parameters. In fact, one can employ the slow-roll 
approximation to effectively align the proposed model 
with the observational data. Hence,
under the slow-roll conditions 
$\vert\ddot \phi\vert\ll \vert3H \dot \phi\vert$ and $\epsilon_1, 
\epsilon_2, \epsilon_3, \ell_0,  \ell_1 \ll1$, 
Eqs. \eqref{2.3}-\eqref{2.5} take the forms 
\begin{equation}
\label{2.9}
3H^2-V(\phi)\simeq 0 ,
\end{equation}
\begin{equation}
\label{2.10}
2\dot H+\dot\phi^2\simeq-6H^2J
(\phi)\dot\phi^2=-2\ell_0H^2,
\end{equation}
\begin{equation}
\label{2.11}
\dot\phi+\frac{V^{\prime}(\phi)}
{3H}\simeq-6H^2J(\phi)\dot\phi=-\frac
{2\ell_0H^2}{\dot\phi}.
\end{equation}

Using Eq. \eqref{2.8}, we have $\epsilon_1 \equiv 
\epsilon_V$ and $\eta \approx -2 \left( \frac{V''}{V} - 
2\epsilon_V \right) = -2\eta_V + 4\epsilon_V$ \cite{41}. 
The slow-roll analysis reveals the significant 
parameters as
\begin{equation}
\label{2.12}
\epsilon_{\rm V}=\frac{1}{2\big(1+
2V(\phi)J(\phi)\big)}\bigg
(\frac{V^\prime(\phi)}{V(\phi)}\bigg)^2,
\end{equation}
\begin{eqnarray}
\label{2.13}
\eta_{\rm V} =&-& \frac{2}{\big(1+2V(\phi)
J(\phi)\big)}\frac{V^{\prime\prime}}{V}+\frac{3}
{2\big(1+2V(\phi)J(\phi)\big)}\bigg
(\frac{V^\prime(\phi)}{V(\phi)}\bigg)^2J(\phi)
\nonumber\\
&+&\frac{4}{\big(1+2V(\phi)
J(\phi)\big)^2}\bigg(\frac{V^\prime(\phi)}
{V(\phi)}\bigg)\bigg(J(\phi)V^\prime(\phi)+
V(\phi)J^\prime(\phi)\bigg).
\end{eqnarray}

Within the framework of the inflationary model scenarios, 
the scalar field value $\phi_e$ at the end of the 
expansion is obtained through  
$\epsilon_1(\phi_e)=1$. Besides, the scalar field 
at the commencement of the inflation can be 
determined by analyzing the total logarithmic 
phase. It is important to note that the number 
of e-folds, related to the inflationary period, is 
crucial for interpreting the cosmological observables. 
It is generally given by 
\begin{equation}
\label{2.14}
N=\int_{t_i}^{t_e}H{\rm d}t=\int_{\phi_i}
^{\phi_e}\frac{H}{\dot\phi}{\rm d}\phi=
-\int_{\phi_i}^{\phi_e} \bigg
(\frac{3H^2+18H^4J(\phi)}{V^\prime(\phi)}\bigg)
{\rm d}\phi.
\end{equation}
In super-horizon scale ($k\ll aH$), the power spectra
of the scalar perturbations demonstrate 
the following $k$-dependence
\begin{equation}
\label{2.15}
P_{\zeta}=\frac{k^3}{2\pi^2}\vert\zeta_k\vert^2
\propto k^{3-2\mu_s},
\end{equation}
where in the mentioned model, the range 
of the scalar disturbances is
\begin{equation}
\label{2.16}
\zeta_k\propto k^{-\mu_s},\;\;\;
\mu_s=\epsilon_1+\frac{1}{2}\eta+\frac{3}{2}.
\end{equation} 
Here, $\zeta_k$ denotes the curvature perturbation mode. 
Therefore, the scalar spectral index becomes
\begin{equation}
\label{2.17}
n_s=1+\frac{d \ln P_{\zeta}}{d\ln k}
=4-2\mu_s=1-2\left(\epsilon_1+\frac{1}{2}\eta\right)
=1 - 6\epsilon_V + 2\eta_V.
\end{equation}
The slow-roll parameter $\ell_0$, which pertains 
to the kinetic coupling, does not appear in 
the scalar spectral index.

An essential quantity is the relative 
contribution of the tensor and
scalar perturbations to the power 
spectra. It is defined via the 
tensor-to-scalar ratio $r$,
\begin{equation}
\label{2.18}
r=\frac{P_T(k)}{P_\zeta(k)}.
\end{equation}
By applying Eq. \eqref{2.15}, the power spectra for 
the scalar perturbations is
\begin{equation}
\label{2.19}
P_{\zeta}=\frac{H^2A_S}{4\pi^2}
\frac{\Bigg[3\big(1-J(\phi)
\dot\phi^2\big)+\frac{3}{V(\phi)
}\Big(\frac{1}{2}\dot
\phi^2-V(\phi)+6V(\phi)J(\phi)
\dot\phi^2\Big)\Big
(\frac{1-J(\phi)\dot\phi^2}{1-3J
(\phi)\dot\phi^2}\Big)
^2\Bigg]^{1/2}}{\Bigg[\frac{1}
{a}\frac{d}{dt}\Big(\frac
{3\dot a}{V(\phi)}\frac{\big(1-J
(\phi)\dot\phi^2\big)
^2}{1-3J(\phi)\dot\phi^2}\Big)-
\Big(1+J(\phi)\dot\phi^2\Big)\Bigg]^{3/2}},
\end{equation}
where Eq. \eqref{2.19} has been derived in 
\cite{32}, in which the perturbations for the 
non-minimal kinetic couplings have been computed 
by using the Sasaki-Mukhanov variable. Also, the 
parameter $A_S$ is defined as follows
\begin{equation}
\label{2.20}
A_S=4^{\mu_s-2}\Bigg\vert\frac{\Gamma
(\mu_s)}{\Gamma(3/2)}\Bigg\vert^2.
\end{equation}
The evaluation of all magnitudes occurs at the moment 
of the horizon exit, in which $k=a H$. Analogously, 
the power spectra for the tensor perturbations have 
the feature \cite{32},
\begin{equation}
\label{2.21}
P_T=16\frac{H^2A_T}{4\pi^2}\frac{\Big[1-J(\phi)\dot\phi^2
\Big]^{1/2}}{\Big[1+J(\phi)\dot\phi^2\Big]^{3/2}},
\end{equation}
\begin{equation}
\label{2.22}
A_T=4^{\mu_T-2}\Bigg\vert\frac{\Gamma
(\mu_T)}{\Gamma(3/2)}\Bigg\vert^2.
\end{equation}
In the limit $\epsilon_1, \epsilon_2, \epsilon_3, 
\ell_0, \ell_1\ll1$, and using  
$\mu_s=\epsilon_1+\frac{1}{2}\eta+\frac{3}{2}$ and 
$\mu_T=\epsilon_1+\frac{3}{2}$, we obtain $A_T=A_S$.
By considering the $P_\zeta$ and $P_T$ in
Eqs. \eqref{2.19} and \eqref{2.21}, up to the 
first order, and using the condition $\epsilon_1, 
\epsilon_2, \epsilon_3, \ell_0, \ell_1\ll1$, the 
parameter $r$ is approximately given by 
\begin{equation}
\label{2.23}
r=16\epsilon.
\end{equation}

In order to compare the results with the 
experimental data, one can also investigate 
the running of the scalar spectral index
$\alpha_s$, which is defined as
\begin{equation}
\label{2.24}
\alpha_s=\frac{dn_s}{d\ln k}=\frac{d^2\ln 
P_{\zeta}}{d\ln k^2},
\end{equation}
where $d/d\ln k=-d/dN$ \cite{42}. According 
to the Planck 2018 data, the parameters 
of the cosmic disturbances must meet the 
following observational limitations \cite{27}-\cite{29},
\begin{equation}
\label{2.25}
n_s=0.9663\pm0.0041,\qquad r<0.065 , 
\qquad \alpha_s=-0.0041\pm0.0067 \tag{2.25},
\end{equation}
with $68\%$ C.L and $95\%$ C.L, respectively.

\section{The dynamical inflation}
\label{300}

The DI represents a distinct class of 
inflationary models that emerge in the early universe. 
This framework employs dynamic processes, including non-
linear field dynamics, phase transitions, 
and moduli stabilization, 
to drive inflation. Such models specifically aim to 
overcome limitations of conventional inflationary 
scenarios by stabilizing scalar fields (potential 
inflaton candidates) and utilizing dynamical 
mechanisms to produce an appropriately flat 
potential \cite{37}-\cite{42}.

The DI relies on the dynamical mechanisms to generate 
the inflationary potential. These mechanisms often 
involve the non-perturbative effects, such as the gaugino 
condensation or the moduli stabilization 
in the string theory, 
which naturally give rise to some flat potentials suitable 
for inflation. An appropriate DI potential is \cite{38},
\begin{eqnarray}
\label{3.1}
V(\phi)=V_0\bigg[1+\bigg(\frac{\phi}{\mu}
\bigg)^{-p}\bigg],
\end{eqnarray}
\begin{equation}
\label{3.2}
V_0=M^4 ,  \qquad  \mu^p=\frac{\Lambda_
3^{p+4}}{M^4},
\end{equation}
where ``$p$'' is a free parameter, and 
the scale of $\Lambda_3$ is determined by 
the associated gauge group. 

We apply the DI potential \eqref{3.1}, and we also 
consider the following non-minimum kinetic coupling 
$J(\phi)$,
\begin{equation}
\label{3.3}
J(\phi)=\frac{1}{J_0}\bigg[\alpha+\beta\bigg
(\frac{\phi}{\mu}\bigg)^{-p}\bigg],
\end{equation}
here we assume that $V_0=J_0$, and also $\alpha$ 
and $\beta$ are arbitrary parameters. 
In the action \eqref{2.1}, if 
$J(\phi)=0$ we can obtain the slow-roll parameters.
Thus, the spectral index $n_s$ can
be articulated through the slow-roll parameters as
\begin{equation}
\label{3.4}
n_s=1-2\epsilon_1-\eta=1 - 6\epsilon_V + 2\eta_V.
\end{equation}

The running of the scalar spectral 
index is given by \cite{41},\cite{42},
\begin{equation}
\label{3.5}
\alpha_s=-2\xi^2-8\epsilon_{\rm V}^2+10 
\epsilon_{\rm V}\eta_{\rm V},
\end{equation}

The higher-order slow-roll parameter 
$\xi^2$, which contributes to the 
running $\alpha_s$, is defined as
\begin{eqnarray}
\label{3.6}
\xi^2 &\equiv& \frac{M_{\rm Pl}^4}{16\pi^2}
\frac{H^{\prime}(\phi)H^{\prime\prime
\prime}(\phi)}{H^2(\phi)}
\nonumber\\
&\simeq&\frac{M_{\rm Pl}^4}{64\pi^2}
\Bigg[\frac{V^{\prime}(\phi)V^{\prime
\prime\prime}(\phi)}{V^2(\phi)}-\frac{3}{2}
\bigg(\frac{V^{\prime}(\phi)}{V(\phi)}\bigg)^2
\frac{V^{\prime\prime}(\phi)}{V(\phi)}
+\frac{3}{4}\bigg(\frac{V^{\prime
\prime}(\phi)}{V(\phi)}\bigg)^4\Bigg].
\end{eqnarray}
This is another parameter, associated with the slow-roll 
dynamics. By utilizing \eqref{2.23}, the  
Eq. \eqref{3.5} can be approximately rewritten as
\begin{equation}
\label{3.7}
\alpha_s \simeq0.12r^2-0.57r(1-n_s)-\frac
{M_{\rm Pl}^4}{32\pi^2}\frac{V^{\prime}(\phi)V^{\prime
\prime\prime}(\phi)}{V^2(\phi)}.
\end{equation}
In fact, the last term in this equation is
paramount in DI models.

The first slow-roll parameter is given by
\begin{equation}
\label{3.8}
\epsilon_{\rm V}(\phi)=\frac{M_{\rm Pl}^2}{2}\bigg(\frac
{V^{\prime}(\phi)}{V(\phi)}\bigg)^2=\frac{p^2 M^2_{\rm Pl}}
{2 \mu^2}\frac{\left(\frac{\mu}{\phi}\right)^{2p+2}}
{\bigg[1+\big(\frac{\mu}{\phi}\big)^p\bigg]^2}\;.
\end{equation}
At the end of the inflation there is 
$\epsilon_{\rm V} \left(\phi_{\rm end}\right) \approx 1$.
Thus, we receive the following 
algebraic equation for the value 
of the inflaton for terminating the inflation 
\begin{equation}
\label{3.100}
\frac{p M_{\rm Pl}}{4\mu}
\left(\frac{\mu}{\phi_{\rm end}}\right)^{p+1}
-\left(\frac{\mu}{\phi_{\rm end}}\right)^p -1 =0.
\end{equation}
For $p=1$, the inflaton $\phi_{\rm end}$ explicitly 
finds the value
\begin{equation}
\phi_{\rm end}=\frac{M_{\rm Pl}}{2\left(
1+\sqrt{1+M_{\rm Pl}/\mu}\right)}.
\end{equation}
If $\mu\gg M_{\rm Pl}$, this obviously reduces to
$\phi_{\rm end} \approx M_{\rm Pl}/4$.
For general value of $p$ and also the case 
$\phi_{\rm end}\ll\mu$, the denominator of 
Eq. \eqref{3.8} tends to $(\mu/\phi)^{2p}$
(or equivalently the third
term of Eq. \eqref{3.100} becomes negligible).
Hence, Eq. \eqref{3.100} possesses the solution  
\begin{equation}
\phi_{\rm end}\approx\frac{p M_{\rm Pl}}{4} .
\end{equation}
This value of the inflaton is independent of the 
parameter $\mu$, while it is proportional to 
the Planck mass. Note that the 
condition $\phi_{\rm end}\ll\mu$ 
clearly defines large value for the parameter $\mu$, i.e.,
$\mu\gg \frac{p}{4} M_{\rm Pl}$.

If $\phi_{\rm end}\gg \mu$, the denominator of 
Eq. \eqref{3.8} tends to 1 (or equivalently the second 
term of Eq. \eqref{3.100} becomes negligible).
In this case, the reduced form of Eq. \eqref{3.100} gives
\begin{equation}
\phi_{\rm end} \approx \left(\frac{p \mu^p M_{\rm Pl}}{4}
\right)^{\frac{1}{p+1}}.
\end{equation}
Hence, the inequality $\phi_{\rm end}\gg \mu$
finds the feature $\mu \ll p M_{\rm Pl}/4$.
Now we transform the parameter $\mu$ to another 
parameter $\gamma$, i.e., 
$\mu=\gamma M_{\rm Pl}$. Thus,, the above $\phi_{\rm end}$
takes the form 
\begin{equation}
\phi_{\rm end} \approx \left(\frac{p \gamma^p}{4}
\right)^{\frac{1}{p+1}} M_{\rm Pl}.
\end{equation}
Therefore, the inequality $\phi_{\rm end}\gg \mu$
imposes the following condition on the parameter
$\gamma$, i.e., $\gamma \ll p/4$.

We observe that by adjusting the parameters we receive  
various values for $\phi_{\rm end}$, which  
are proportional to the Planck mass $M_{\rm Pl}$. 
This result prominently is consistent with the 
literature \cite{27}, \cite{43}-\cite{49}.

The second slow-roll parameter is
\begin{equation}
\label{3.10}
\eta_{\rm V}(\phi)=\epsilon_{\rm V}-\frac
{M_{\rm Pl}}{4\sqrt\pi}\frac
{\epsilon_{\rm V}^{\prime}}{\sqrt\epsilon_{\rm V}}
=\bigg(\frac{\phi_0}{\phi}\bigg)^{2(p+1)}+\frac
{p+1}{2\sqrt\pi}\frac{M_{\rm Pl}}{\phi}\bigg(\frac
{\phi_0}{\phi}\bigg)^{p+1},
\end{equation}
where $\phi_0=M_{\rm Pl}\bigg(\frac
{p}{4\sqrt\pi}\lambda\bigg)^{1/(p+1)}$. In the case 
$\lambda \ll (\phi/M_{\rm Pl})^p$, with 
$\lambda = (\mu/M_{\rm Pl})^p$, the constant $V_0$ in
Eq. \eqref{3.1} plays a dominant role in the potential.  

It is crucial to note that when 
$\phi\simeq\phi_0\ll M_{\rm Pl}$, 
the parameter $\eta_{\rm V}$ becomes very large
and $\epsilon_{\rm V}(\phi) \simeq 1$, which suggest a 
breakdown of the slow-roll approximation. In 
other words, it is inconsistent 
to the state that the inflation 
initiates at $\phi=\phi_0$.   
For the second case $\phi\gg\phi_0$, 
both $\epsilon_{\rm V}$ and $\eta_{\rm V}$ become small, 
which admits a consistent slow-roll approximation. 
In the third case 
$\phi_0\ll\phi\ll M_{\rm Pl}$, the second term 
of $\eqref{3.10}$ becomes the dominant 
factor, which leads to the condition 
$\eta_{\rm V}\gg\epsilon_{\rm V}$. 
Thus, $\eta$ can be reformulated as in the following 
\begin{equation}
\label{3.11}
\eta_{\rm V}(\phi)\simeq\frac
{p+1}{2\sqrt\pi}\sqrt{\epsilon_{\rm V}(\phi)}\frac
{M_{\rm Pl}}{\phi}=\frac
{\lambda p(p+1)}{8\pi}\bigg(\frac
{M_{\rm Pl}}{\phi}\bigg)^{p+2}.
\end{equation}

The number of e-folds can be expressed as
\begin{eqnarray}
\label{3.12}
N&=&\frac{2\sqrt\pi}{M_{\rm Pl}}
\int^{\phi_c}_{\phi}\frac{d\phi^{\prime}}{\sqrt
{\epsilon_{\rm V}(\phi^{\prime})}}
\simeq \frac{p+1}{p+2}\bigg(\frac{1}{\eta_{\rm V}
(\phi_c)}-\frac{1}{\eta_{\rm V}(\phi)}
\bigg), \quad \epsilon_{\rm V}\ll \eta_{\rm V} .
\end{eqnarray}
The field value $\phi_c$ refers 
to the critical point in which
inflation has ceased. We shall 
consider $\phi_c$ as a free parameter. 
However, combining Eqs. \eqref{3.11} and \eqref{3.12},
and also imposing $\phi \ll \phi_c$, 
the number of e-folds $N$ tends to the following 
constant value
\begin{equation}
\label{3.13}
N_{\rm tot}\equiv \frac{p+1}{p+2}
\frac{1}{\eta_{\rm V}(\phi_c)}=\frac{8\pi}{\lambda p(p+2)}
\bigg(\frac{\phi_c}{M_{\rm Pl}}\bigg)^{p+2}.
\end{equation}
This upper limit can be very large. By defining 
$\phi_N$ as the field value at the $N$ e-folds before 
inflation, we can articulate $\eta_{\rm V} (\phi_N)$ 
in terms of $N$ and $N_{tot}$,
\begin{equation}
\label{3.14}
\eta_{\rm V} (\phi_N)=\frac{p+1}{p+2}\frac{1}{N_{tot}-N}.
\end{equation}
Consequently, when $N\ll N_{\rm tot}$, $\eta_{\rm V}$ 
approaches to a constant value. 
The spectral index can be written as
\begin{equation}
\label{3.15}
n_s=1+2\eta_{\rm V}-6\epsilon_{\rm V}
\simeq 1+\frac{p+1}{p+2}
\frac{2}{N_{\rm tot}}\bigg(1
-\frac{50}{N_{\rm tot}}\bigg)^{-1}.
\end{equation}
Thus, the running of the scalar spectral index can be 
represented as a function of the spectral index 
\begin{equation}
\label{3.16}
\alpha_s =-\frac{1}{2}\frac{p+2}{p+1}(n_s-1)^2.
\end{equation}

In the special case where $J(\phi)=0$, 
the values for $n_s$ and $r$ 
approximately fall within the 
ranges $n_s\in[1, 1.06]$ 
and $r\in[10^{-14}, 10^{-7}]$ 
for $p=2$; $n_s\in[1, 1.06]$ 
and $r\in[10^{-17}, 10^{-6}]$ 
for $p=3$; and $n_s\in[1, 1.06]$ and $r\in[10^
{-20}, 10^{-4}]$ for $p=4$. The values of $r$ 
precisely are consistent with the Planck data, 
while the values of $n_s$ do not 
align with the same dataset \cite{50}.

Now consider the case where $J(\phi)$ takes 
the form of \eqref{3.3}.
Thus, the slow-roll parameters and the other related 
factors are modified. In order to calculate $n_s$ 
and $r$, it is useful to employ the parameters  
$\epsilon_{\rm V}$ and $\eta_{\rm V}$. They can 
be represented as follows
\begin{eqnarray}
\label{3.17}
\epsilon_{\rm V}&=&\frac{1}{4}\frac{ p^2
\big(\frac{\phi}{\mu}\big)^{-2p}
}{{\phi}^{2}\left( 1+ \big( {\frac {\phi}
{\mu}} \big) ^{-p} \right) ^{2} \Bigg
[\beta\big(\frac{\phi}{\mu}\big)^{-2p}+
\left( \beta+\alpha \right)  \big( {\frac
{\phi}{\mu}}\big) ^{-p}+\alpha+
\frac{1}{2} \Bigg]}\;,
\end{eqnarray}
\begin{eqnarray}
\label{3.18}
\eta_{\rm V}&=&\frac{7\Bigg[\beta\big(\frac{\phi}{\mu}\big)
^{-2p}+ \left( \beta+\alpha \right) 
\big( {\frac {\phi}{\mu}}\big) ^{-p}+\alpha+
\frac{1}{2}\Bigg]^{-2}}{{4\phi}^{2} \left( 1+ 
\big( {\frac {\phi}{\mu}} \big) ^{-p} \right) 
^{2}}
\nonumber\\
&\times&\Bigg\{p \big( {\frac {\phi}{\mu}}
\big)^{-p}\Bigg[\beta\big( p-\frac{4}{7} \big)
\big(\frac{\phi}{\mu}\big)^{-3p}+ \left
(\alpha\big(\frac{3}{7}\,p-\frac{4}{7}\big)+
\beta\big( p-{\frac {8}{7}} \big)\right)
\big(\frac{\phi}{\mu}\big)^{-2p}
\nonumber\\
&+&\left(\alpha\big( -\frac{p}{7}-
{\frac {8}{7}}\big)
-\frac{p}{14}-\frac{4}{7}\,\beta-\frac{2}{7}
\right)  \big( {\frac {\phi}{\mu}} \big) ^{-p}-
\frac{4}{7}\big( \alpha+\frac{1}{2} \big)
\left( p+1 \right)\Bigg]\Bigg\}.
\end{eqnarray}

The number of e-folds has the form 
\begin{eqnarray}
\label{3.19}
N&=&\frac{\phi^2}{p( p-1)( p-2)( p+2)\left( {\frac {\phi}
{\mu}} \right)^{-p}}\Bigg\{\big( -\beta\,{p}^{2}+
4\,\beta\big)\big( {\frac {\phi}{\mu}}
\big)^{-3p}
\nonumber\\
&-&4( p-1)( p+2)\big(\frac{1}{2}\alpha+\beta\big)
\big( {\frac {\phi}{\mu}}\big)^{-2p}+
( p-1)( p-2)( p+2)\big(\beta+2\,\alpha+\frac{1}{2}\big)
\big({\frac {\phi}{\mu}} \big)^{-p}\nonumber\\
&+&2( p-1)( p-2)( \alpha+\frac{1}{2}) \Bigg\}.
\end{eqnarray}

To obtain the observables as functions 
of $N$, we numerically 
invert Eq. \eqref{3.19} for $\phi(N)$ and substitute 
into Eqs. \eqref{3.16}-\eqref{3.18}. This is done for each 
parameter set, which ensures consistency with the 
slow-roll approximation. In the same 
way, both the tensor-to-scalar ratio and 
the spectral index takes the features  
\begin{eqnarray}
\label{3.20}
r&=&\frac{8p^2\big( {\frac {\phi}{\mu}}
\big)^{-2p}}{{\phi}^{2} \left
( 1+ \big( {\frac{\phi}{\mu}} \big) ^{-p} \right) ^
{2} \Bigg[1+2\left(1+\big( {\frac {\phi}{\mu}} \big)
^{-p}\right)\left( \alpha+\beta\big( {\frac {\phi}
{\mu}} \big)^{-p}
\right)\Bigg]}\;,
\end{eqnarray}
\begin{eqnarray}
\label{3.21}
n_s&=&1+\frac{1}{ {\phi}^{2}\left( 1+\big
( {\frac {\phi}{\mu}}\big) ^{-p} \right)
\Bigg[\beta\big( {\frac {\phi}{\mu}}
\big)^{-2p}+ \left( \beta+\alpha \right)  \left
( {\frac {\phi}{\mu}}\right) ^{-p}+\alpha+
\frac{1}{2} \Bigg]}
\nonumber\\
&\times&\Bigg\{ p \left( p+1 \right)\big( {\frac
{\phi}{\mu}} \big) ^{-p} -\frac{{p}^{2}
\left( 2\,\beta\, \big( {\frac {\phi}{\mu}}
\big) ^{-p}+\beta+\alpha \right)\big( {\frac {\phi}{\mu}}
\big)^{-2p} 
}{\beta\big( {\frac {\phi}{\mu}}\big)^{-2p}+ \left( \beta+
\alpha \right)  \left( {\frac {\phi}{\mu}}
\right) ^{-p}+\alpha+\frac{1}{2}}
\nonumber\\
&-&\frac{5}{4}\frac{{p}^{2}\big( {\frac {\phi}{\mu}}
\big)^{-2p}}{\left( 1+\big( {\frac {\phi}{\mu}
}\big) ^{-p} \right)}\Bigg\}.
\end{eqnarray}
The running spectral index also finds the form
\begin{eqnarray}
\label{3.22}
\alpha_s&=&-\frac{9}{4{\phi}^
{4} \left(1+\big({\frac
{\phi}{\mu}}\big)^{-p} \right) ^{4}
\Bigg[\beta\big( {\frac {\phi}{\mu}}
\big)^{-2p}+\left
(\beta+\alpha \right)\left( {
\frac {\phi}{\mu}} \right)^{-p}
+\frac{1}{2}+\alpha \Bigg]^{4}}
\nonumber\\
&\times&\Bigg\{{p}^{2}\big( {\frac {\phi}{\mu}}
\big)^{-2p}\Bigg\{
\Bigg[{\beta}^{2}\left( p-1 \right)
\big( p-\frac{4}{9} \big)
\big( {\frac {\phi}{\mu}}
\big)^{-6p}
\nonumber\\
&+&\frac{3}{2}\beta\, \left( \big( \beta+
\frac{5}{9}\,\alpha \big)
{p}^{2}+ \big(-{\frac {26\,\beta}{9}}-{\frac
{40\,\alpha}{27}} \big) p+{\frac {32
\,\beta}{27}}+{\frac {16\,\alpha}
{27}}\right)\big( {\frac {\phi}{\mu}}
\big)^{-5p}
\nonumber\\
&+&\Bigg(\big( -{\frac {13\,
\alpha\,\beta}{18}}
+{\frac {5\,{\alpha}^{2}}{18}}
-{\frac {7\,\beta}{18}} \big) 
{p}^{2}+ \big( -6\,
\alpha\,\beta-{\frac {7\,{\alpha}
^{2}}{9}}-\frac{13}{3}\,
{\beta}^{2}-{\frac {7\,\beta}{9}}\big) p
\nonumber\\
&+&\frac{4}{9}\,{\alpha}^{2}+
\frac{8}{3}\,{\beta}^{2}
+\frac{4}{9}\,\beta+{\frac {32\,\alpha\,
\beta}{9}}\Bigg) \big( {\frac {\phi}{\mu}}
\big)^{-4p}\Bigg]
\nonumber\\
&+&\Bigg[\bigg(-\frac{4}{9}
\,{\alpha}^{2}+ \big(
-{\frac {25\,\beta}{6}}-\frac{1}
{12} \big) \alpha
-\frac{1}{2}\,{\beta}^{2}-{\frac 
{61\,\beta}{36}}\bigg)p^2
\nonumber\\
&+&\left( -\frac{5}{3}\,{\alpha}
^{2}+ \big( -\frac{14}{3}
\,\beta-\frac{4}{9}\big) \alpha-
{\frac {13\,{\beta}^
{2}}{9}}-{\frac {11\,\beta}{9}}
\right) p+{\frac {16\,{\alpha}^{2}}
{9}}+ \big( \frac{4}{9}+\frac{16}{3}
\,\beta \big) \alpha
\nonumber\\
&+&{\frac {16\,{\beta}^{2}}{9}}+\frac{4}{3}\,\beta
\Bigg] \big( {\frac {\phi}{\mu}}
\big)^{-3p}
+\Bigg[-\frac {17}{6}\big(\frac{1}{2}+\alpha
\big)\big(\frac{9}{17}\alpha+\beta\big)p^2
\nonumber\\
&-&\frac{2}{9}\, \big( \frac{3}
{2}\,\alpha+\beta+\frac{1}{4} \big)
\big( \frac{1}{2}+\alpha \big) p+
\frac{8}{3}\,{\alpha}^{2}+
\big( \frac{4}{3}+{\frac {32\,\beta}
{9}} \big) \alpha
\nonumber\\
&+&\frac{1}{9}+\frac{4}{3}\,\beta+\frac{4}{9}\,{
\beta}^{2}\Bigg]\big( {\frac {\phi}{\mu}}
\big)^{-2p}-\frac{2}{9}\Bigg[\big( \beta+
\frac{5}{2}\,\alpha+\frac{3}{4} \big)p-4\beta
\nonumber\\
&-&8\alpha-2)\Bigg]\left( p+1
\right)\big( \frac{1}{2}+
\alpha\big)  \big( {\frac {\phi}{\mu}}
\big)^{-p}+\frac{2}
{9} \left( p+1 \right)\left( p+2\right) \big(
\frac{1}{2}+\alpha\big)^{2}\Bigg\}\Bigg\}.
\end{eqnarray}

By calculating the parameters $r$, $n_s$ and $\alpha_s$, 
and selecting the values for the free parameters 
$\mu$, $\alpha$ and $\beta$ 
within a specific range, we can 
compare the outcomes of this 
model against the data provided by the Planck 2018. 
Taking $\alpha=\beta=0$, these quantities 
reduce to the standard gravity model
with the coupling constant $\kappa$, as expected 
\cite{51}, \cite{52}.

In the Tables~\ref{tab:hresult1}--\ref{tab:hresult3},
by assigning specific values to the free parameters,
we obtained results for the significant parameters 
(slow-roll parameters) of our model, 
namely $r$ and $n_s$, together with the running of 
the spectral index $\alpha_s$. These 
results are entirely consistent with the Planck 
data. In contrast, for the case $J(\phi)=0$, as 
previously discussed, the results do not correspond 
closely with the Planck data, 
except for the parameter $r$.

\newpage
\begin{table}[h]
\caption{The numerical calculations for the 
tensor-to-scalar ratio $r$, scalar spectral index $n_s$, 
and the running spectral index $\alpha_s$ have been  
performed over $N=50 \to 60$ e-folds, via the 
parameters $\alpha=-0.02$, $\beta=10$, $p=3$, 
and different values of $\mu$. } 
\centering 
\begin{tabular}{c rrrr} 
\hline\hline 
$N$&$\mu$& $n_s\quad\quad\quad$& $
r\quad\quad\quad$&
$\alpha_s\quad\quad\quad$  \\ [0.5ex]
\hline 
50$\ \rightarrow$\ 60 & 2.0 & 0.9627$ 
\ \rightarrow$\ 0.9485 &\ 15$\times 10^{-4}$$ 
\rightarrow$\ 2.0$\times 10^{-3}
$& -0.00101$\rightarrow
$\ -0.00197\\
50$\ \rightarrow$\ 60 & 2.1 & 0.9700$ 
\ \rightarrow$\ 0.9615 &\ 12$
\times 10^{-4}$$\rightarrow
$\  1.6$\times 10^{-3} $& -0.00065$
\rightarrow$\ -0.00109\\ 
50$\ \rightarrow$\ 60 & 2.2 &
\ 0.9750$ 
\ \rightarrow$\ 0.9695 &\ 10$
\times 10^{-4}$$\rightarrow
$\  1.2$\times 10^{-3}$& -0.00044$
\rightarrow$\ -0.00067\\
50\ $ \rightarrow$\ 60 & 2.3 
&\ 0.9788$\ 
\rightarrow$\ 0.9749&\ 8.0$
\times 10^{-4}$$\rightarrow
$\  1.0$\times 10^{-3}$& -0.00032$
\rightarrow$\ -0.00045\\
50\ $ \rightarrow$\ 60 & 2.4 & 0.9817$\ 
\rightarrow$\ 0.9789 &\ 7.0$\times 10
^{-4}$$\rightarrow
$\  8.0$\times 10^{-4}$& -0.00024$
\rightarrow$\ -0.00032\\
[1ex] 
\hline 
\end{tabular}
\label{tab:hresult1}
\end{table}

\begin{table}[h]
\caption{The numerical calculations for the 
tensor-to-scalar ratio $r$, scalar spectral index 
$n_s$, and the running spectral index $\alpha_s$ 
have been performed over $N=50 \to 60$ e-folds, 
via the parameters $\mu=2$, $\beta=10$, $p=3$, 
and different values of $\alpha$} 
\centering 
\begin{tabular}{c rrrr} 
\hline\hline 
$N$&$\alpha$& $n_s\quad\quad
\quad$& $r\quad\quad\quad$
&$\alpha_s\quad\quad\quad$  \\ [0.5ex]
\hline 
50\ $ \rightarrow$\ 60 & -0.07 & 0.9659
$
\ \rightarrow$\ 0.9543 &\ 12.9$\times 10
^{-4}$$\rightarrow
$\  1.7$\times 10^{-3}$& -0.00085$
\rightarrow$\ -0.00155\\
50\ $ \rightarrow$\ 60 & -0.09 & 0.9672$\ 
\rightarrow$\ 0.9566 &\ 11.8$
\times 10^{-4}$$\rightarrow
$\  1.6$\times 10^{-3}$& -0.00079$
\rightarrow$\ -0.00140\\
50\ $ \rightarrow$\ 60 & -0.11 & 0.9685$\ 
\rightarrow$\ 0.9589 &\ 10.8$
\times 10^{-4}$$\rightarrow
$\  1.4$\times 10^{-3}$& -0.00072$
\rightarrow$\ -0.00125\\
50\ $ \rightarrow$\ 60 & -0.13 & 0.9698$\ 
\rightarrow$\ 0.9611 &\ 9.0$\times 10^
{-4}$$\rightarrow
$\  1.2$\times 10^{-3}$& -0.00066$
\rightarrow$\ -0.00112\\
50\ $ \rightarrow$\ 60 & -0.15 & 0.9712$\ 
\rightarrow$\ 0.9634 &\ 8.0$\times 10^
{-4}$$\rightarrow
$\  1.1$\times 10^{-3}$& -0.00060$
\rightarrow$\ -0.00099\\
50\ $ \rightarrow$\ 60 & -0.17 & 0.9725$\ 
\rightarrow$\ 0.9656 &\ 7.9$\times 10^
{-4}$$\rightarrow
$\  1.0$\times 10^{-3}$& -0.00055$
\rightarrow$\ -0.00088\\
50\ $ \rightarrow$\ 60 & -0.19 & 0.9739$\ 
\rightarrow$\ 0.9677 &\ 7.1$\times 10
^{-4}$$\rightarrow
$\  0.8$\times 10^{-3}$& -0.00049$
\rightarrow$\ -0.00077\\
50\ $ \rightarrow$\ 60 & -0.21 & 0.9753$\ 
\rightarrow$\ 0.9699 &\ 6.0$\times 
10^{-4}$$\rightarrow
$\  0.7$\times 10^{-3}$& -0.00044$
\rightarrow$\ -0.00067\\
[1ex] 
\hline 
\end{tabular}
\label{tab:hresult2}
\end{table} 

\begin{table}[h]
\caption{The numerical calculations for the 
tensor-to-scalar ratio $r$, scalar spectral index 
$n_s$, and the running spectral index $\alpha_s$ 
have been performed over $N=50 \to 60$ e-folds, 
via the parameters $\alpha=-0.02$, $\mu=2$, $p=3$, 
and different values of $\mu$.} 
\centering 
\begin{tabular}{c rrrr} 
\hline\hline 
$N$&$\beta$& $n_s\quad\quad\quad$& $r\quad\quad
\quad$&$\alpha_s\quad\quad
\quad$  \\ [0.5ex]
\hline 
50$\ \rightarrow$\ 60 & 10.4 & 0.9671$ 
\ \rightarrow$\ 0.9566 &\ 13.0$
\times 10^{-4}$$ 
\rightarrow$\ 1.8$\times 10^{-3}
$& -0.00078$\rightarrow$\ 
-0.00139\\
50$\ \rightarrow$\ 60 & 10.5 & 0.9681$ 
\ \rightarrow$\ 0.9583 &\ 12.7$
\times 10^{-4}$$\rightarrow
$\  1.7$\times 10^{-3} $& -0.00074$
\rightarrow$\ -0.00129\\ 
50$\ \rightarrow$\ 60 & 10.6 &\ 0.9690$ 
\ \rightarrow$\ 0.9598 &\ 12.0$\times 
10^{-4}$$\rightarrow
$\  1.6$\times 10^{-3}$& -0.00070$
\rightarrow$\ -0.00119\\
50\ $ \rightarrow$\ 60 & 10.7 &\ 0.9698$\ 
\rightarrow$\ 0.9612&\ 11.8$\times 
10^{-4}$$\rightarrow
$\  1.5$\times 10^{-3}$& -0.00066$
\rightarrow$\ -0.00111\\
50\ $ \rightarrow$\ 60 & 10.8 & 0.9706$\ 
\rightarrow$\ 0.9626 &\ 11.3$
\times 10^{-4}$$\rightarrow
$\  1.4$\times 10^{-3}$& -0.00062$
\rightarrow$\ -0.00103\\
50\ $ \rightarrow$\ 60 & 10.9 & 0.9714$\ 
\rightarrow$\ 0.9638 &\ 10.9$
\times 10^{-4}$$\rightarrow
$\  1.4$\times 10^{-3}$& -0.00059$
\rightarrow$\ -0.00096\\
50\ $ \rightarrow$\ 60 & 11.0 
& 0.9722$\ 
\rightarrow$\ 0.9650 &\ 10.5$
\times 10^{-4}$$\rightarrow
$\  1.3$\times 10^{-3}$& -0.00056$
\rightarrow$\ -0.00090\\
50\ $ \rightarrow$\ 60 & 11.1 
& 0.9729$\ 
\rightarrow$\ 0.9661 &\ 10.1$
\times 10^{-4}$$\rightarrow
$\  1.2$\times 10^{-3}$& -0.00053$
\rightarrow$\ -0.00084\\
[1ex] 
\hline 
\end{tabular}
\label{tab:hresult3}
\end{table}

After comparing the results in these tables
for $r$, $n_s$ and $\alpha_s$   
with Eq. $\eqref{2.25}$, we observe that the 
results are in excellent agreement with 
the Planck 2018 data.

It is clear that the parameters $\alpha$, $\beta$ 
and $\mu$ must be fixed within a specific range. 
Hence, we do not have  
freedom to choose any values for them. Therefore, 
this is a very good result that our free parameters 
must be in a very specific range to be consistent 
with the Planck data. These results, which are stated 
in the above tables, have been plotted 
in the Figures \ref{fig:CMB1}--\ref{fig:CMB3}. 

In the Figure \ref{fig:CMB1}, we have shown the results from 
the data in the table~\ref{tab:hresult1}. In this case, 
$\alpha$ and $\beta$ have 
specific fixed values, while $\mu$ is variable, which  
changes in the interval $\mu\in [2,\ 2.4]$. It can 
be seen that the results of the 
$n_s$ - $r$ curve are in good 
agreement with the Planck data. Thus, we can claim 
that our model covers a large part of the 
experimental data, i.e., the Planck 2018 and 
BICEP/Keck dataset along with the recently 
released CMB-S4, and Stage 3 \cite{53}, \cite{54}.
It follows from the Figure \ref{fig:CMB1} that $r$ decreases with 
increasing $\mu$ and fixing $\alpha$ 
and $\beta$. By taking large 
values of $\mu$, the value of $n_s$ decreases, 
and also $n_s$ increases with increasing $\mu$. 

\subsection{Comparison with the recent 
ACT and CMB-S4 results}

The Atacama Cosmology Telescope (ACT) DR6 
\cite{55}, \cite{56} 
data and the latest CMB-S4 forecasts further restrict the 
tensor-to-scalar ratio to $r < 0.035$ (95 \% C.L.) 
and slightly tighten the bounds on 
$n_s = 0.9655\pm0.0035$. The predictions 
of our model for $r\sim10^{-3}$--$10^{-2}$ 
and $n_s\simeq0.95$--$0.98$ 
remain well within these new limits. We 
have checked that the 
running $\alpha_s$ stays in the range 
$|\alpha_s|<10^{-3}$, which is 
compatible with both ACT and Planck results. 
Finally, we also 
compared our results with the latest ACT DR6 data, which 
agrees well with Planck data, and provides deeper primary 
CMB measurements. Our parameter ranges satisfy the 
updated constraints, which confirm the model's viability.
\begin{center}
\begin{figure}
\centering
\includegraphics[width=15cm]{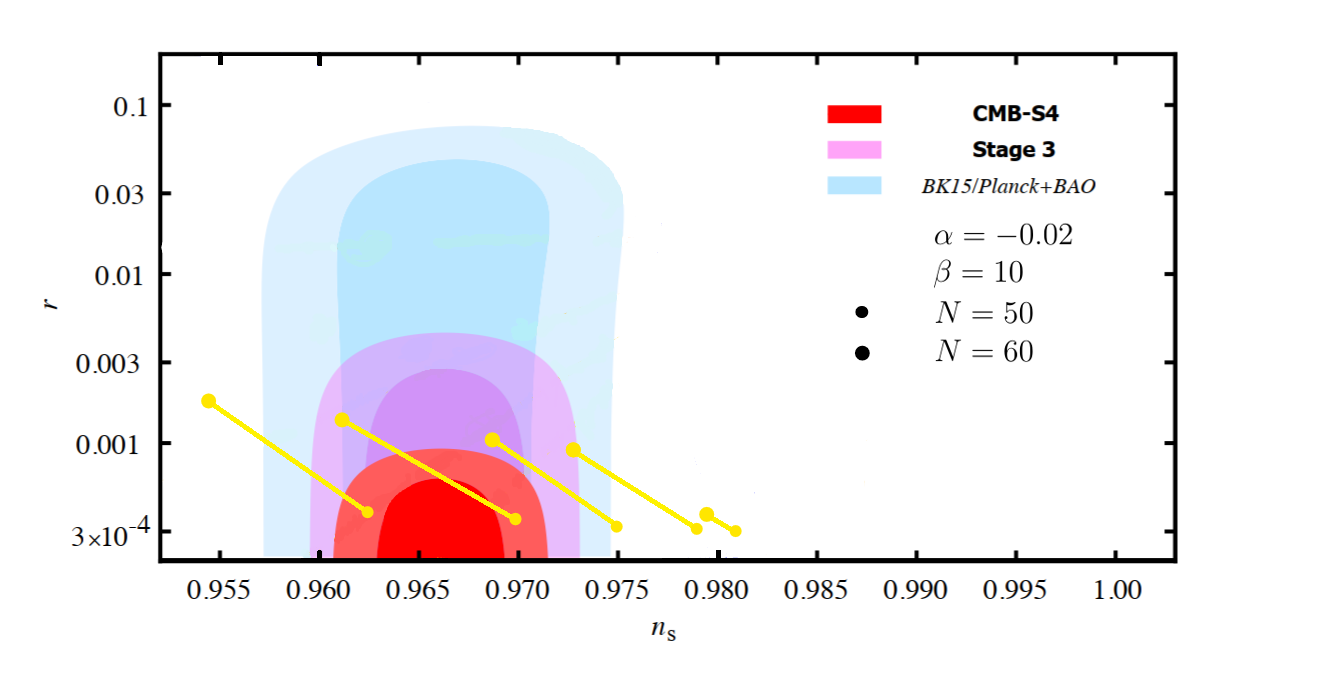}
\caption{\scriptsize{The tensor-to-scalar ratio 
$r$ versus the 
scalar spectral index $n_s$ for $N=50$--$60$ e-folds. 
The solid lines show the model's predictions 
for $\alpha=-0.02$, $\beta=10$, $p=3$, in which 
$\mu$ varies in the range $2.0 \leq \mu \leq 2.4$ 
(from top to bottom of the $r$ values). 
The $1\sigma$ and $2\sigma$ confidence 
contours from the Planck 2018+BICEP/Keck 2018 data 
have been shown for comparison. 
For this parameter range,
the model's predictions fall well within the 
observational constraints.}}
\label{fig:CMB1}
\end{figure}
\end{center}

\begin{center}
\begin{figure}
\centering
\includegraphics[width=15cm]{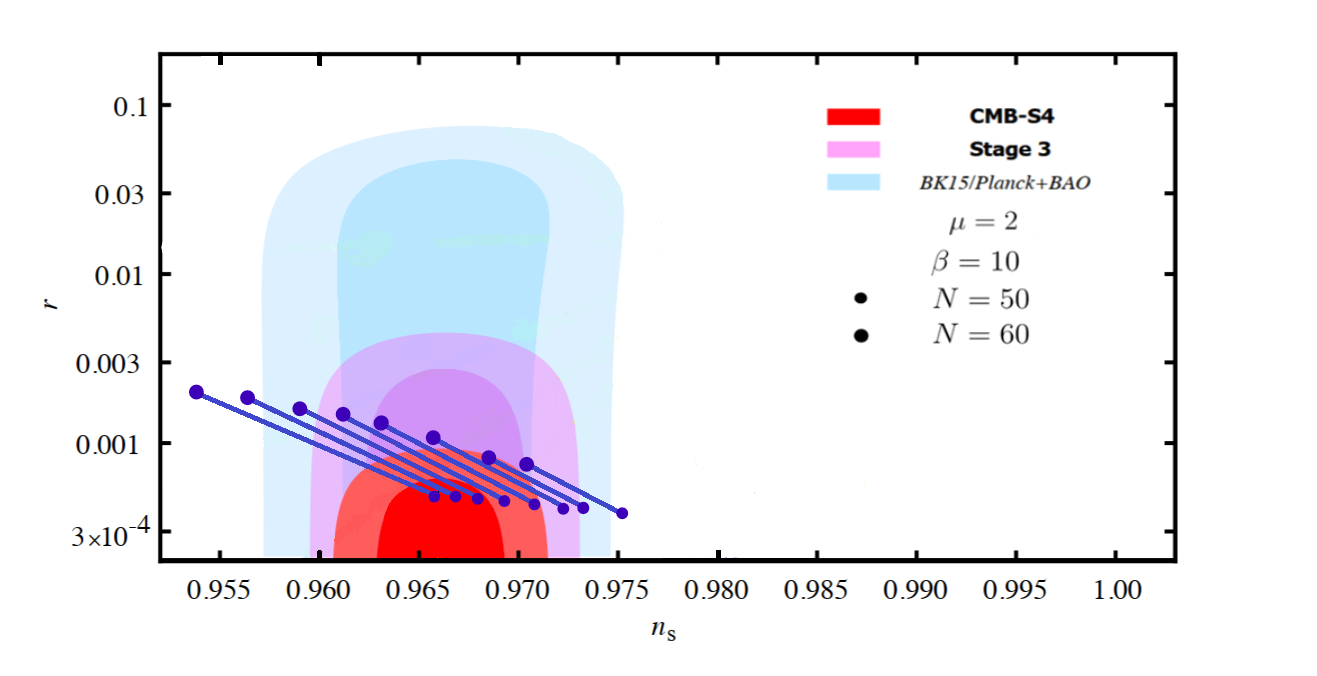}
\caption{\scriptsize{The tensor-to-scalar ratio $r$ versus 
the scalar spectral index $n_s$ for $N=50$--$60$ e-folds. 
The solid lines show the model's predictions for $\beta=10$, 
$\mu=2$, $p=3$, and $\alpha$ varies in the 
range $-0.21 \leq \alpha \leq -0.07$ (from top to 
bottom of the $r$ values). The predictions have been 
plotted against the Planck 2018+BICEP/Keck 2018 confidence 
contours.}}
\label{fig:CMB2}
\end{figure}
\end{center}

Next, we analyze the Table~\ref{tab:hresult2}. 
Its results have been  
plotted in the Figure \ref{fig:CMB2}. 
Here, we consider the 
parameters $\mu$ and $\beta$ to be fixed 
and the parameter $\alpha$ as a variable which  
changes in the interval $\alpha\in[-0.21, -0.07]$. 
We observe that the results of the curve  
$n_s$ - $r$ are in agreement with the Planck data. 
By changing just one of the above  
parameters, we can cover a large portion of the 
data of the Planck satellite, CMB-S4 and Stage 3.
For the Figure \ref{fig:CMB2}, the value of $r$ 
decreases with decreasing 
$\alpha$ and fixing $\beta$ and 
$\mu$. By taking small values for $\alpha$, the 
range of $n_s$ decreases, and $n_s$ increases 
with decreasing $\alpha$.

\begin{center}
\begin{figure}
\centering
\includegraphics[width=15cm]{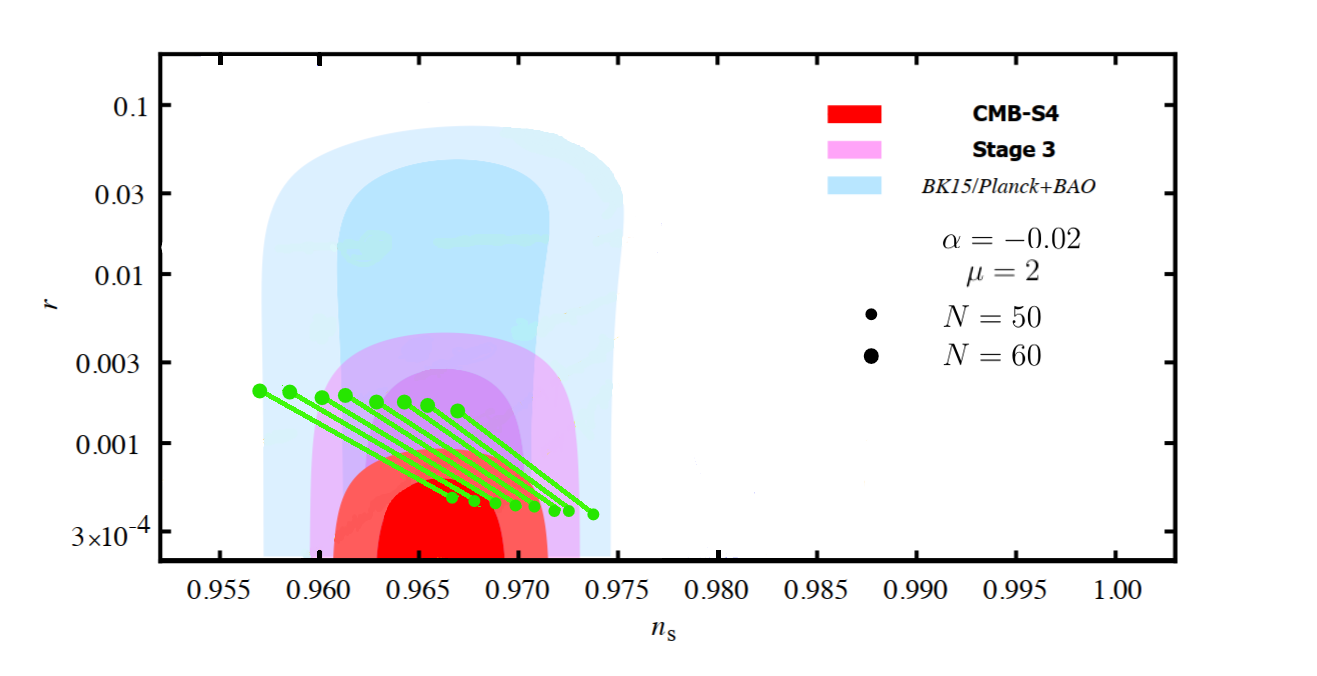}
\caption{\scriptsize{The tensor-to-scalar ratio $r$ 
versus the scalar spectral index $n_s$ for $N=50$--$60$ 
e-folds. The solid lines show the model's predictions 
for $\alpha=-0.02$, $\mu=2$, $p=3$, and $\beta$ 
varies in the range $10.4 \leq \beta \leq 11.1$ 
(from top to bottom of the $r$ values). 
The results are consistent with the observational bounds
of the Planck 2018+BICEP/Keck 2018.}}
\label{fig:CMB3}
\end{figure}
\end{center}

Finally, the results of our 
analysis of the Table~\ref{tab:hresult3} have been  
shown in the Figure \ref{fig:CMB3}. Here, 
we assumed that $\alpha$ 
and $\mu$ are fixed, while $\beta$ is a
variable, which changes in the range 
$\beta \in [10.4, 11.1]$. By              
determining the slow-roll parameters, we can 
compare the experimental data with the most 
suitable and optimal selections for each 
of the three free parameters in each of the three 
tables. This analysis allows us to extract extremely 
favorable outcomes. For the Figure \ref{fig:CMB3}, 
the values of $r$ and $n_s$ decrease with 
increasing $\beta$ and fixing  
$\mu$ and $\alpha$, and the value of 
$n_s$ increases with increasing $\beta$.

As it was shown in the Figures 
\ref{fig:CMB1}--\ref{fig:CMB3}, 
we have presented the correlation between 
the spectral indices $r$ and $n_s$. Then, we 
compared the results, which were obtained 
from the slow-roll parameters, with the 
experimental data. They clearly yielded 
favorable results. It is important to note that 
all these implications and results are in the presence 
of our $J(\phi)$.

These behaviors are plotted in the Figures 
\ref{fig:CMB1}--\ref{fig:CMB3} by 
implementing the Planck line constraints 
\cite{27}, \cite{28}. 
The values of the number of e-folds, 
which are established by the experimental 
values, belong to the interval $N\in[50, 60]$. Thus, the  
behaviors of the curve $n_s-r$ 
with varying $\alpha$, $\beta$, 
and $\mu$ in the mentioned ranges are investigated. 

When the scalar potential and scaled gravity 
are coupled with a kinetic term, it yields 
intriguing numerical values for $n_s$ and $r$. 
These values fall within a favorable range of the 
Planck data, which includes both the 95\% and 
68\% CL contour regions of the recently published 
BICEP/Keck data\cite{27}-\cite{29}. The Planck 
findings, which are smaller than 0.1, are in good 
agreement with the range of $r$, which is 
$[6\times 10^{-4},2\times 10^{-3}]$. 
It has been predicted that a kinetic term 
could be added to the coupling between 
the scalar potential and the scaled gravity 
\cite{52}, \cite{57}. 

\begin{center}
\begin{figure}
\centering
\includegraphics[width=16cm]
{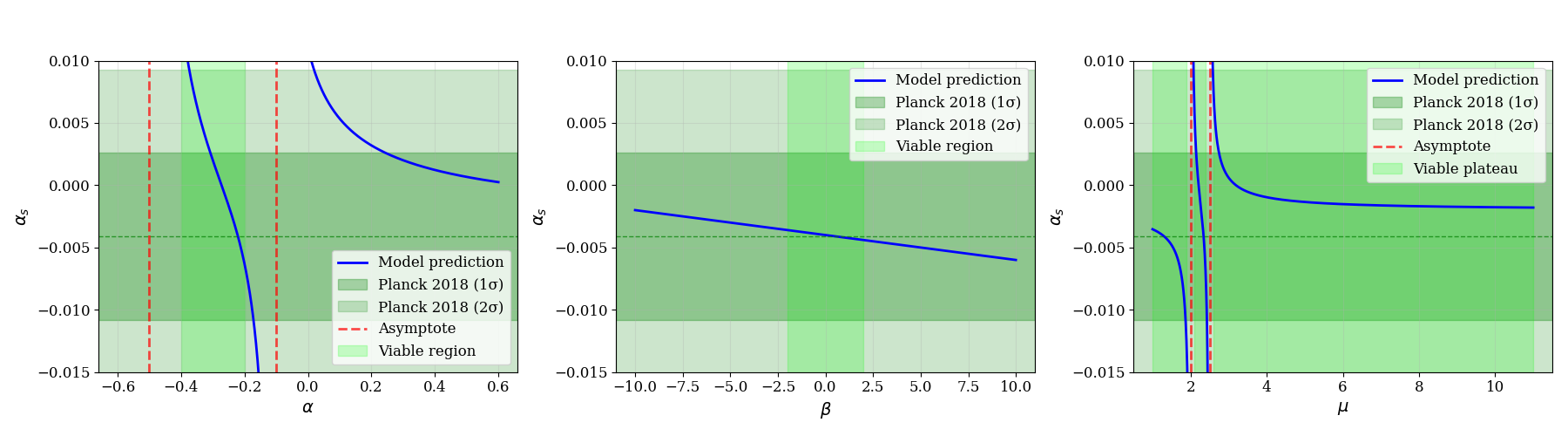}
\caption{\scriptsize{The parameter-space analysis 
of the running of the spectral index $\alpha_s$ 
shows distinct 
functional dependencies. Left: $\alpha_s(\alpha)$ exhibits 
asymptotic divergence at $\alpha = -0.5$ and $-0.1$, which
delineates the theoretical boundaries in the coupling 
parameter-space. Middle: $\alpha_s$ linearly decreases 
with $\beta$, for the moderate $\beta$ values, it
remains within the Planck's $2\sigma$ constraints. 
Right: $\alpha_s(\mu)$ reveals a
resonant structure with the singularities at 
$\mu = 2.0$ and $2.5$, 
creating observationally viable plateaus in the intervening 
parameter regions. The theoretical prediction 
(the blue curve) has been  
evaluated against the observational constraints 
$\alpha_s = -0.0041 \pm 0.0067$ from the Planck 
collaboration.}}
\label{fig:alpha_s_sensitivity}
\end{figure}
\end{center}

\begin{center}
\begin{figure}
\centering
\includegraphics[width=16cm]{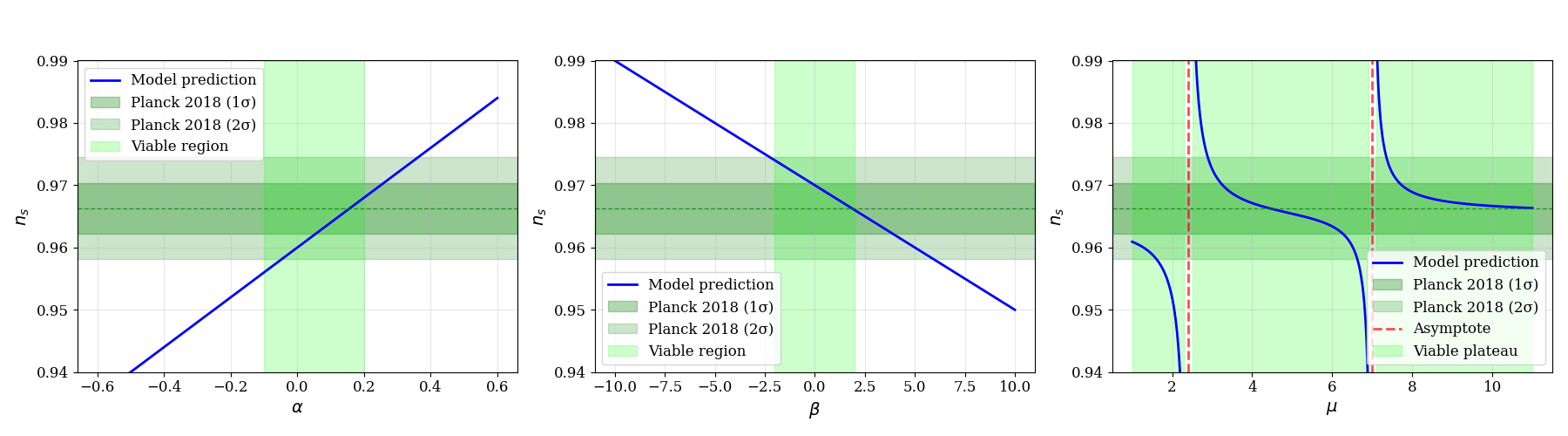}
\caption{\scriptsize{The sensitivity analysis of $n_s$ 
across the fundamental inflationary parameters. 
Left: $n_s(\alpha)$ monotonically increases 
with the coupling 
strength $\alpha$, which intersects the Planck preferred 
value $n_s = 0.9663 \pm 0.0041$ at 
intermediate $\alpha$ values. 
Middle: $n_s(\beta)$ linearly decreases with the field 
parameter $\beta$; the central values of 
$\beta$ yield observationally consistent results. 
Right: $n_s(\mu)$ features, separated by the asymptotes 
at $\mu = 2.4$ and $7.0$, indicate the parameter regions 
where the spectral index 
remains stable against the mass scale variations. 
All viable regions 
(lemon) satisfy the observational constraints 
within the $2\sigma$ confidence levels.}}
\label{fig:n_s_sensitivity}
\end{figure}
\end{center}

\begin{center}
\begin{figure}
\centering
\includegraphics[width=16cm]{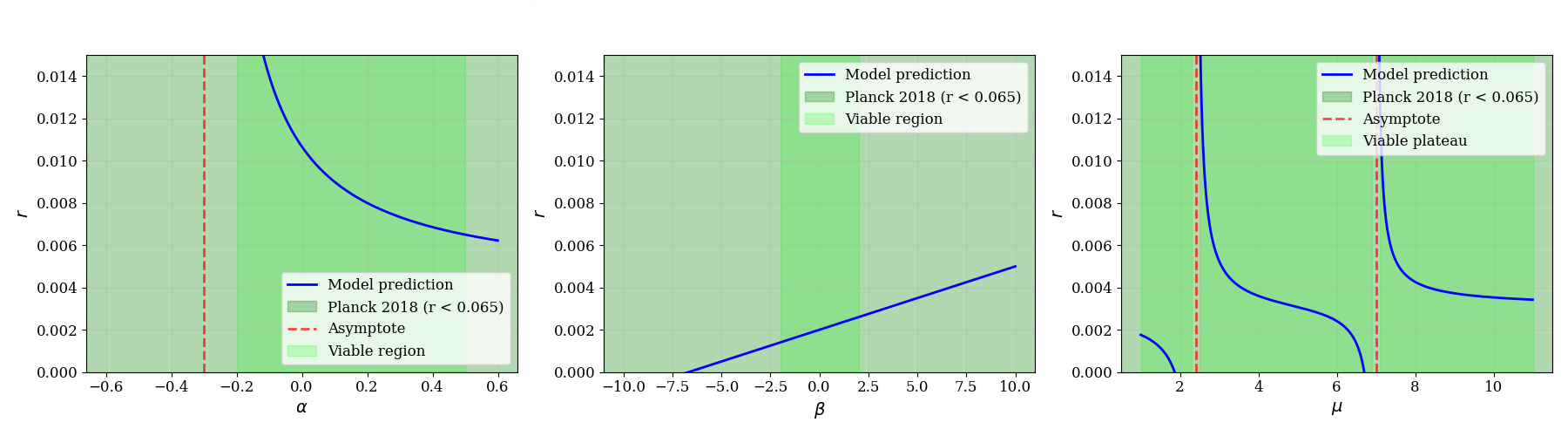}
\caption{\scriptsize{The dependence of $r$ on the 
inflationary model parameters. Left: $r(\alpha)$ 
displays a singular behavior at $\alpha = -0.3$, 
which indicates 
critical enhancement of the tensor mode generation. 
Middle: $r(\beta)$ exhibits linear growth with the  
parameter $\beta$, which remains consistent with the 
Planck upper bound $r < 0.065$ for $\beta \lesssim 20$. 
Right: $r(\mu)$ shows resonant amplification 
near $\mu = 2.4$ and $7.0$. 
The green-shaded region represents the observationally 
permitted parameter-space, where the viable 
inflationary trajectories 
occur when the theoretical predictions fall below the 
experimental limit.}}
\label{fig:r_sensitivity}
\end{figure}
\end{center}

The Figures \ref{fig:alpha_s_sensitivity}--
\ref{fig:r_sensitivity} show the sensitivity 
plots of the parameters $r$, $n_s$ and $\alpha_s$ with 
respect to the parameters $\alpha$, $\beta$ and $\mu$.

\begin{itemize}
    \item Each subplot combines the theoretical predictions 
(blue curves) with the observational constraints 
(green bands) from the Planck 2018 data.
    \item The vertical red-dashed lines indicate the 
theoretical asymptotes where the model predictions diverge.
    \item The colored backgrounds highlight the stable 
parameter regions, in which the lemon indicates the  
observationally viable zones.
\end{itemize}

The sensitivity analysis, presented in Figures 
\ref{fig:alpha_s_sensitivity}--\ref{fig:r_sensitivity}, 
provides crucial 
insight into the viable parameter 
space of our model. The presence 
of the vertical asymptotes 
in these plots indicates the 
values of $\alpha$, $\beta$ and 
$\mu$ where the denominators in the expressions for 
$n_s$, $r$ and $\alpha_s$ vanish, leading to a divergence. 
These asymptotes are unphysical and represent a breakdown 
of the slow-roll approximation or an instability in the 
underlying theory for those specific parameter values. 
Consequently, the physically meaningful and observationally 
viable parameter-space is not located at these asymptotes, 
but rather in the stable regions between them. In these 
plateaus, the cosmological observables attain finite, nearly 
constant values that are robust against small variations in 
the parameters. The ranges for $\alpha$, $\beta$ and $\mu$,
where they have been  
quoted in our analysis and they were 
used to generate the results in 
Tables~\ref{tab:hresult1}--\ref{tab:hresult3} 
and the Figures \ref{fig:CMB1}--\ref{fig:CMB3}, are  
chosen from within these 
stable regions. This demonstrates that the 
agreement with the Planck data is not a fine-tuned 
coincidence at a singular 
point, but is a robust prediction 
of the model across finite 
and well-defined intervals of its parameter-space.

\subsection{Systematic analysis of the parameter 
space and naturalness}
\label{sec:parameter_analysis}

A principal result of this work is 
the identification of the 
regions in the model's parameter-space that yield 
predictions consistent with the latest cosmological 
constraints. To address this systematically, 
we have conducted 
a comprehensive numerical scan over the free 
parameters $\alpha$, $\beta$ and $\mu$, with the 
power $p$ fixed to 3. This is for 
evaluating the resulting spectral 
index $n_s$, the tensor-to-scalar ratio $r$ and the 
running $\alpha_s$ for the cosmologically relevant 
range of $N = 50$ to $60$ e-folds.

The allowed ranges of the parameters, 
as rigorously determined by this scan and are summarized in 
the Tables~\ref{tab:hresult1}--\ref{tab:hresult3}, are:
\begin{itemize}
    \item $\alpha \in [-0.21, -0.07]$
    \item $\beta \in [10.4, 11.1]$
    \item $\mu \in [2.0, 2.4] \, M_{\rm Pl}$
\end{itemize}
These ranges are not isolated points 
but continuous intervals, which 
demonstrate that the agreement with the data is a robust 
prediction of the model and does not rely on the extreme 
fine-tuning. The corresponding observables fall within the 
following windows:
\begin{itemize}
    \item $n_s \in [0.9485, 0.9817]$
    \item $r \in [0.0015, 0.007]$
    \item $\alpha_s \in [-0.00197, -0.00024]$
\end{itemize}

The question of the naturalness is paramount. 
The parameter $\mu$, which is of the order of the Planck 
mass $M_{\rm Pl}$, is a 
natural scale for an effective field 
theory which describes the inflation. 
The parameters $\alpha$ and $\beta$, which govern the 
strength and structure of the non-minimal kinetic coupling, 
are of the order $\mathcal{O}(-0.02)$ and $\mathcal{O}(10)$, 
respectively. Crucially, they are 
not hierarchically large or small. 
That is, we do not require 
$\beta / \alpha \sim 10^{10}$. 
The values are technically natural in the sense that 
setting $\alpha, \beta \to 0$ 
restores the standard GR limit, 
and the observed values are stable under small variations, 
as evidenced by the plateaus in the sensitivity plots 
(Figures~\ref{fig:alpha_s_sensitivity}--
\ref{fig:r_sensitivity}).

Besides, the model itself imposes a theoretical 
constraint that further informs 
the naturalness. As it will be detailed 
in the Conclusions section, 
the conditions $\alpha \neq \pm \beta$ 
emerge from the requirement of a physically consistent 
slow-roll inflation and a graceful exit. 
The viable parameters 
ranges, that we have automatically identified,  
satisfy this theoretical 
prior. Therefore, the observationally allowed parameter 
space is not an arbitrary finely-tuned subset. It is 
meaningfully constrained by both theoretical consistency 
and experimental data. This synergy suggests that the 
presented parameter ranges are not only allowed but they
are also theoretically preferred within 
the framework of our model.

\section{Conclusions}
\label{400}

We investigated the slow-roll dynamics 
corresponding to a scalar-tensor model that  
includes a non-minimal kinetic coupling. The 
potential is the DI and the coupling 
was represented as a power functional of the 
scalar field, characterized by two free 
parameters. In the context of the 
standard canonical scalar field, the DI 
potential yields significant solutions 
that characterize various phases of the cosmological 
development, which includes the 
accelerated expansion in the late universe. 
Besides, it provides solutions for the early-time inflation.

The spectral index $n_s$ and 
$r$ depend on the three free parameters 
$\alpha$, $\beta$ and $\mu$. When these parameters are 
quenched, the resulting values of $n_s$ become 
inconsistent with the observational constraints. These 
parameters can be used to adjust the scalar field values 
at the end and at the beginning of the 
inflation. The $n_s-r$ curves 
have been shown for the different cases. 
The Figures \ref{fig:CMB1}--\ref{fig:CMB3}
were plotted for various values of $\alpha$, $\beta$ 
and $\mu$. Precisely, in each figure, two parameters 
have been fixed while the third parameter changes in a 
specified interval.

The model exhibits two specific limits. The first, when 
$\alpha$ = -\ $\beta$ and here is the $\mu$ in the 
specified range [2, 2.4]. We derive $n_s$ = 0.98140 
but $r$ is incompatible with the Planck constraints.
The second, when $\alpha$ = $\beta$, with a specific 
and defined value for $\mu\in[2, 2.4]$, 
we obtain $r$ = 0.0044, 
but the outcomes of $n_s$ are inconsistent with the 
experimental data. Thus, for the physical consistency,
these constraints necessarily 
impose $\alpha\neq \pm \beta $. 
Across the cosmologically relevant range $N\in$\ [50, 60], 
the model generates observationally viable parameters: 
$r\in$\ [0.0015, 0.007], $n_s\in$\ [0.9485, 0.9817] and 
$\alpha_s\in$\ [-0.00197, -0.00024]. The aforementioned 
results are consistent with the $\alpha\neq \pm \beta$ 
conditions, otherwise, the slow rolling condition 
during the inflation and at the 
end of it would be violated.

The sensitivity diagrams of the quantities $r$, $n_s$ and 
$\alpha_s$ versus the free parameters has been displayed. 
Our explanation revealed that the 
free parameters cannot take 
any arbitrary range. We found that 
$n_s$, $r$ and $\alpha_s$ 
fall within the ranges that have been 
established by the most 
recent observational evidence. 
Actually, our sensitivity analysis 
reveals that the theoretical parameter-space is naturally 
segmented by asymptotes. The 
requirement for the cosmological 
observables $n_s$, $r$ and $\alpha_s$ 
to be finite and stable 
directly restricts the parameters 
$\alpha$, $\beta$ and $\mu$ 
to the plateaus between these asymptotic boundaries. The 
observationally viable ranges, that we have 
identified, are not arbitrarily 
chosen but they are a direct 
consequence of this underlying 
structure. The latter one inherently 
prevents the parameters 
from taking on unphysical values that 
would lead to divergences.



\begin{thebibliography}{99}

\bibitem{1}
A. H. Guth, Phys. Rev. {\textbf {D 23}} (1981) 
347-356.

\bibitem{2}
A. D. Linde, Phys. Lett. {\textbf {B 108}} (1982) 393.

\bibitem{3}
H.P Nilles, Phys. Rep. {\textbf {110}} (1984) 1-162.

\bibitem{4}
E. Witten, Nucl. Phys. {\textbf {B 188}} (1981) 513-554.

\bibitem{5}
D. Baumann, L. McAllister, ``\emph{Inflation 
and String Theory. Cambridge University
Press}'', 2015.

\bibitem{6}
K.I Izawa, and T. Yanagida, Theor. Phys. 
{\textbf {95}} (1996) 829-830.

\bibitem{7}
M. Arai, S. Kawai and N.Okada, Phys. 
Lett. {\textbf {B 650}} (2007) 133-138.

\bibitem{8}
K. Nakayama, F. Takahashi and T.Watari, 
JHEP {\textbf {12}} (2011) 031.

\bibitem{9}
A. H. Guth and P.J. Steinhardt, 
Scientific American {\textbf {250}} (1984) 129.

\bibitem{10}
A. D. Linde, JETP Lett. {\textbf {40}} (1984) 1333.

\bibitem{11}
A. Starobinsky, JETP Lett. {\textbf {73}} (2001) 371.

\bibitem{12}
J. A. Adams, B. Cresswell and R. Easther, 
Phys. Rev. {\textbf {D 64}} (2001) 123514.

\bibitem{13}
S. Nojiri, S. D. Odintsov and M. Sami, 
Phys. Rev. {\textbf {D 74}} (2006) 046004.

\bibitem{14}
T. J Li, J. L. Lopez and D. V. Nanopoulos, 
Phys. Rev. {\textbf {D 56}} (1997) 2606.

\bibitem{15}
M. Sami, N. Savchenko and A. Toporensky, 
Phys. Rev. {\textbf {D 70}} (2004) 123528.

\bibitem{16}
X. Chen, R. Easther and E. A. Lim, 
JCAP {\textbf {06}} (2007) 023.

\bibitem{17}
V. K. Oikonomou, 
Class. Quant. Grav. {\textbf {38}} (2021) 195025,

\bibitem{18}
E. D. Stewart, Phys. Rev. {\textbf {D 51}} (1995) 684.

\bibitem{19}
H. V. Peiris, D. Baumann, B. Friedman and A. Cooray, 
Phys. Rev. {\textbf {D 76}} (2007) 103517.

\bibitem{20}
A. G. Cadavid and A. E. Romano, 
Eur. Phys. J. C {\textbf {75}} (2015) 589.

\bibitem{21}
B. Li and J. D. Barrow, 
Phys. Rev. {\textbf {D 75}} (2007) 084010.

\bibitem{22}
A. W. Beckwith, 
AIP Conf. Proc. {\textbf {880}} (2007) 1180.

\bibitem{23}
V. F. Mukhanov and G. V. Chibisov, 
JETP Lett.  {\textbf {33}} (1981) 532.

\bibitem{24}
A. A. Starobinsky and H. J. Schmidt, 
Class. Quant. Grav. {\textbf {4}} (1987) 695.

\bibitem{25}
J. M. Bardeen, 
Phys. Rev. {\textbf {D 22}} (1980) 1882. 

\bibitem{26}
J. M. Bardeen, P. J. Steinhardt and M. S. Turner, 
Phys. Rev. {\textbf {D 28}} (1983) 679.

\bibitem{27}
Y. Akrami et al. ``\emph{Planck 2018 results. 
X. Constraints on inflation}'', Astron. Astrophys.
{\textbf {641}} (2020) 10, 
[arXiv:1807.06211 [astro-ph.CO]].

\bibitem{28}
N. Aghanim et al. ``\emph{Planck 2018 
results. VI. Cosmological parameters}'', 
Astron. Astrophys. {\textbf {641}} (2020) 6, 
[arXiv:1807.06209 [astro-ph.CO]].

\bibitem{29}
P. A. R. Ade et al. ``\emph{Improved 
Constraints on Primordial Gravitational 
Waves using Planck, WMAP, and BICEP/Keck 
Observations through the 2018 Observing 
Season}'', Phys. Rev. Lett. {\textbf {127}} 
(2021) 151301, [arXiv: 2110.00483 
[astro-ph.CO]].

\bibitem{30}
V. K. Oikonomou,
Phys. Rev. {\textbf {D 103}} (2021) 044036;
B. Li and J. D. Barrow,
Phys. Rev. {\textbf {D 75}} (2007) 084010;
S. Nojiri and S. D. Odintsov, 
Int. J. Geom. Meth. Mod. Phys. {\textbf {4}} 
(2007) 115; T. P. Sotiriou, V. Faraoni, 
Rev. Mod. Phys. {\textbf {82}} (2010) 451;
A. De Felice, S. Tsujikawa, 
Living Rev. Rel. {\textbf {13}} (2010) 3.

\bibitem{31}
K. Bamba, S. Nojiri, S. D. Odintsov and 
D. S$\acute{a}$ez-G$\acute{o}$mez, 
Phys. Rev. {\textbf {D 90}} (1983) 679;
S. Tsujikawa, Lect. Notes 
Phys. {\textbf {800}} (2010) 99;
S. Nojiri, S. D. Odintsov, 
Phys. Rept. {\textbf {505}} (2011) 59-144;
S. Nojiri, S. D. Odintsov, V. K. Oikonomou, 
Phys. Rept. {\textbf {692}} (2017) 1-104.

\bibitem{32}
H. Tiberiu, S. N. L. Francisco, N. 
Shin$\acute{i}$chi and S.D. Odintsov,
Phys. Rev. {\textbf {D 84}} (2011) 024020; 
C. Germani and K. Kehagias,
Phys. Rev. Lett.  {\textbf {105}} (2010)
011302.

\bibitem{33}
P. H. R. S. Moraes, J. D. V. Arba$\tilde{n}$il, 
and M. Malheiro, JCAP {\textbf {06}} (2016) 005.

\bibitem{34}
F. Younesizadeh and D. Kamani, Gen. Rel.  
Grav. {\textbf 57} (2025) 114,
https://doi.org/10.1007/s10714-025-03448-4, 
arXiv:2507.23321 [gr-qc];
F. Younesizadeh and D. Kamani, 
Int. J. Theor. Phys. {\textbf {64}} (2025) 60,
arXiv:2507.15812 [gr-qc];  J. Astrophys. Astr.  
{\textbf {46}} (2025) 42, arXiv:2507.19005 [gr-qc].

\bibitem{35}
F. G. Alvarenga, A. dela Cruz-Dombriz, M. J. S. 
Houndjo, M. E. Rodrigues, and D. S$\acute{a}$ez-
G$\acute{o}$mez, 
Phys. Rev. {\textbf {D 87}} (2013) 10.

\bibitem{36}
S. Bhattacharjee, J. R. L. Santos, P. H. R. S. 
Moraes and P. K. Sahoo,
Eur. Phys. J. Plus {\textbf {137}} (2020) 576.

\bibitem{37}
W.H. Kinney and A. Riotto, 
Astropart. Phys. {\textbf {10}} (1999) 387.

\bibitem{38}
W.H. Kinney and A. Riotto,
Phys. Lett. {\textbf { B 435}} (1998) 272. 
K. Hamaguchi, KL. Izawa, and H. Nakajima. 
Physics Letters {\textbf { B 662}} (2008) 208-12.
I. Affleck, M. Dine and N. Seiberg. 
Nucl.Phys. {\textbf {B 256}} (1985) 557.
K. Schmitz and T. T. Yanagida.
Phys. Rev. {\textbf {D 94}} (2016) 074021.

\bibitem{39}
G. W. Horndeski,
Int. J. Theor. Phys. {\textbf{10}} (1974) 363--384.

\bibitem{40}
T. Kobayashi, M. Yamaguchi, and J. Yokoyama,
Prog. Theor. Phys. {\textbf{126}} (2011) 511--529.

\bibitem{41}
D. J. Schwarz, C. A. Terrero-Escalante and A. A. Garca,
Phys. Lett.  {\textbf {B 517}} (2001) 243; 
S. M. Leach et al. 
Phys. Rev. {\textbf {D 66}} (2002) 023515.

\bibitem{42}
 A. Kosowski and M.S. Turner, 
Phys. Rev. {\textbf { D 52}} (1995) 1739.

\bibitem{43}
 J. Martin and C. Ringeval, 
JCAP {\textbf {08}} (2006) 009.

\bibitem{44}
A. Linde, Phy. Rept. {\textbf {333-334}} (2000) 575-591.

\bibitem{45}
D. Baumann, [arXiv preprint, arXiv:0907.5424].

\bibitem{46}
L. Kofman, A. Linde, and A.A. Starobinsky,
Phys. Rev. {\textbf {D 56}} (1997) 3258.

\bibitem{47}
K. D. Lozanov and M. A. Amin,
Phys. Rev. {\textbf {D 90}}. 2014, 083528.

\bibitem{48}
L. Boubekeur and D. H. Lyth,
JCAP {\textbf {07}} (2005) 010.

\bibitem{49}
D. H. Lyth and A. Riotto,
Phy. Rept. {\textbf {314}} (1999) 1-146.

\bibitem{50}
J. Martin, C. Ringeval and V. Vennin.
``\emph{Encyclopaedia Inflationaris}'', 
Phys. Dark Univ. 5-6 (2014) 75-235
[arXiv: arXiv:1303.3787 [astro-ph.CO]].

\bibitem{51}
N. A. Avdeev and A. V. Toporensky,
Gravitation and Cosmology {\textbf {28}} 
(2022) 416; Q. Huang, H. Huang, F. Tu, 
L. Zhang and J. Chen,
Annals Phys. {\textbf {409}} (2019) 
167921; L. Wu, Q. Gao, Y. Gong, Y. Jia and T. Li,
Commun. Theor. Phys.  {\textbf {73}} (2021) 
075402; S. Teymourtashlou and D. Kamani, Eur. Phys. J.
{\bf C 81} (2021) 761, arXiv:2108.10164 [hep-th]; 
D. Kamani, Phys. Lett. {\bf B 564} (2003) 123-131,
arXiv:hep-th/0304236 [hep-th].

\bibitem{52}
N. A. Avdeev and A. V. Toporensky,
Gravitation and Cosmology {\textbf {27}} (2021) 269;
I. V. Fomin, S. V. Chervon
and A. V. Tsyganov, Eur. Phys. 
J. C {\textbf {80}} (2020) 350;
R. Fakir and W. G. Unruh,
Phys. Rev. {\textbf {D 41}} (1990) 1783-1791;
Z. K. Guo and D. J. Schwarz, Phys. 
Rev. {\textbf {D 81}} (2010) 123520;
H. Daniali and D. Kamani, Nucl. Phys. {\bf B 975}
(2022) 115683, arXiv:2202.09347 [hep-th].

\bibitem{53}
Sebastian Belkner et al. ``\emph{CMB-S4: 
Iterative Internal Delensing and r Constraints
}'', Astrophys. J. 964 (2024) 2, 148, 
[arXiv:  2310.06729 [astro-ph.CO]].

\bibitem{54}
M. D. Niemack, ``\emph{Designs for a large-
aperture telescope to map the CMB 10x 
faster,}'', Appl. Opt. {\bf 55} (2016) 1688-1696,
[arXiv:1511.04506 [astro-ph.IM]].

\bibitem{55}
T. Louis et al., ``Atacama Cosmology 
Telescope DR6: Updated 
Constraints on Inflationary Parameters,'' 
[arXiv:2503.14454 [astro-ph.CO]] (2025).

\bibitem{56}
E. Calabrese et al., ``ACT DR6: 
Cosmological Parameters with 
Planck and BAO,'' [arXiv:2503.14452 
[astro-ph.CO]] (2025).

\bibitem{57}
S. Tsujikawa, Phys. Rev. {\textbf { D 85}} (2012) 083518.

\end{thebibliography}
\end{document}